\documentclass[
reprint,
 aps,
 superscriptaddress,
bibnotes,
pra
]{revtex4-2}
\usepackage{physics}
\usepackage{dsfont}
\usepackage{color,newfloat}
\usepackage{graphicx}
\usepackage{dcolumn}
\usepackage{bm}
\usepackage{amsmath,amssymb,amsthm}
\usepackage{mathtools}
\usepackage{multirow}
\usepackage{hhline}
\usepackage{comment}
\usepackage[dvipsnames]{xcolor}
\usepackage{hyperref}
\hypersetup{
    colorlinks=true,
    linkcolor=Maroon,
    citecolor=Maroon,
    urlcolor=Maroon
}

\usepackage{gensymb} 
\usepackage[caption = false]{subfig}
\usepackage{lineno}
\usepackage{siunitx}
\usepackage[mathscr]{euscript}
 \let\mathscr\relax
\usepackage{makecell}

\begin{document}

\title{
A programmable quantum photonic platform integrating coherent emitters 
}

\author{Hanna Salamon}
\thanks{These authors contributed equally.}
\affiliation{Center for Hybrid Quantum Networks (Hy-Q), Niels Bohr Institute, University of Copenhagen, Jagtvej 155A, 2200 Copenhagen, Denmark}

\author{Mikkel T. Mikkelsen}
\thanks{These authors contributed equally.}
\affiliation{Center for Hybrid Quantum Networks (Hy-Q), Niels Bohr Institute, University of Copenhagen, Jagtvej 155A, 2200 Copenhagen, Denmark}

\thanks{These authors contributed equally.}

\author{Zhe Liu}
\affiliation{Center for Hybrid Quantum Networks (Hy-Q), Niels Bohr Institute, University of Copenhagen, Jagtvej 155A, 2200 Copenhagen, Denmark}

\author{Peter Lodahl}
\affiliation{Center for Hybrid Quantum Networks (Hy-Q), Niels Bohr Institute, University of Copenhagen, Jagtvej 155A, 2200 Copenhagen, Denmark}

\author{Severin Kr{\"u}ger}
\affiliation{Experimental Physics VI,, Ruhr-Universit{\"a}t Bochum, Universit{\"a}tsstrasse 150, D-44780 Bochum, Germany}

\author{Arne Ludwig}
\affiliation{Experimental Physics VI,, Ruhr-Universit{\"a}t Bochum, Universit{\"a}tsstrasse 150, D-44780 Bochum, Germany}

\author{Leonardo~Midolo}
\email{midolo@nbi.ku.dk}
\affiliation{Center for Hybrid Quantum Networks (Hy-Q), Niels Bohr Institute, University of Copenhagen, Jagtvej 155A, 2200 Copenhagen, Denmark}


\begin{abstract}
Photonic quantum technologies require the integration of high-quality quantum light sources with programmable photonic circuitry to enable on-chip quantum information processing. Semiconductor quantum dots have emerged as a leading platform for quantum light generation, yet their monolithic integration with cryogenic-compatible, programmable photonic circuitry remains challenging. Here we demonstrate a programmable quantum photonic platform that enables simultaneous control of highly coherent quantum dot-based single photon emitters with reconfigurable optical elements at cryogenic temperatures. System-level integration enables individual electrical and optical addressing of active components on a single chip, allowing their concurrent operation while preserving coherent non-classical light emission. We demonstrate multiple quantum-photonic functionalities, including pure single-photon generation, two-photon interference, and resonant transmission and reflection measurements of integrated emitters. These results mark a step from isolated quantum-optical devices towards programmable quantum-photonic hardware platforms capable of combining coherent quantum emitters with large-scale photonic functionality.
\end{abstract}


\date{\today}

\maketitle

\section{Introduction} 
The long-term vision of quantum photonics is the realization of integrated circuits in which quantum states are generated, processed, and distributed with the level of functionality and programmability established by modern electronic processors.
Recent advances in integrated photonic platforms, including silicon photonics and thin-film lithium niobate, have enabled increasingly complex and programmable optical circuits capable of implementing routing {\cite{carolan2015universal, bogaerts2020programmable,paesani2019generation}}, interference {\cite{politi2008silica, crespi2013integrated, silverstone2014chip, kuttner2026heralded}}, and high-speed switching {\cite{wang2018integrated, zhu2021integrated}} for both classical and quantum information processing. These developments have driven a transition from individual components to large-scale reconfigurable photonic processors. 
In parallel, semiconductor quantum dots (QDs) in III--V materials have demonstrated state-of-the-art performance in terms of brightness, single-photon purity, and indistinguishability {\cite{uppu2020scalable,tomm2021bright, kuhlmann2015transform, loredo2026deterministic}}, establishing themselves as a leading solid-state platform for deterministic photon-emitter interfaces, single- and entangled-photon generation, and nonlinear processing \cite{lodahl2015interfacing, le2021experimental, javadi2015single, meng2024deterministic}.\\
However, these two technological trajectories have largely evolved independently. Platforms enabling large-scale programmability generally lack mature coherent quantum emitters, while platforms capable of generating state-of-the-art quantum light have yet to demonstrate comparable levels of circuit complexity and active hardware integration. Bridging this divide is one of the central challenges facing integrated quantum photonics. 
The realization of a common platform that combines coherent quantum emitters with scalable control and reconfigurable photonic functionality remains an outstanding milestone, which goes beyond the demonstration of individual components. \\
Scaling quantum communication and computing hardware requires the orchestration of many indistinguishable emitter-photon interfaces and optical elements to overcome the imperfections, disorder, and  inhomogeneity arising from self-assembly and nanofabrication. \\
In III--V quantum photonics, the challenge lies mainly in developing reconfigurable circuitry compatible with spectrally-tunable QDs. The cryogenic environment precludes thermo-optic approaches to switching and routing, while the realization of efficient electro-optic circuitry in GaAs is hindered by the weak electro-optic effect and electroabsorption. An emerging solution is offered by nano-electromechanical systems (NEMS) \cite{midolo2018nano, errando2019mems}, offering compact, energy-efficient, and intrinsically low-loss photonic elements that are compatible with cryogenic operation. To date, NEMS have been extensively employed for phase shifting, strain tuning, and photon routing, in both III--V and silicon quantum photonics \cite{papon2019nanomechanical, qvotrup2025integration, Buzzi2025SpectralStrain, larocque2024tunable, bishop2018electro, midolo2012spontaneous}, but their integration with coherent quantum emitters has not been demonstrated.\\
In this work, we developed a cryogenic GaAs quantum-photonic circuit that combines multiple independently controlled active components within a single architecture. The device integrates two individually tunable p-i-n-doped QD emitters with a reconfigurable NEMS-based photon router and an electrical control infrastructure that enables independent operation of the different subsystems. We further show that QDs retain high single-photon purity and narrow emission linewidths approaching the Fourier-transform limit, even after repeated warm-up and cooldown cycles.

\begin{figure*}[!ht]
    \centering
    \includegraphics{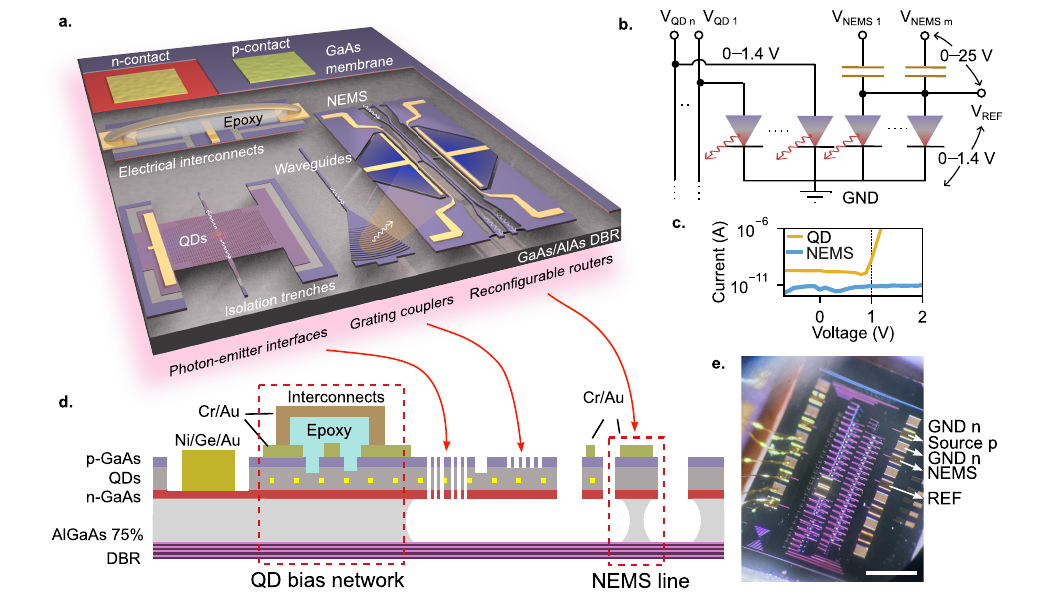}
    \caption{\textbf{Programmable GaAs platform for quantum photonics.} \textbf{(a.)} Illustration of the integrated platform based on GaAs p-i-n membrane and the individual components. \textbf{(b.)} Schematic of the electrical interconnects to simultaneously drive QDs and NEMS. The NEMS reference voltage (REF) is used to suppress the diode built-in field across the circuit. \textbf{(c.)} Measured current-voltage characteristics for QD diode and NEMS. \textbf{(d.)} Optical stack and processing steps. \textbf{(e.)} Photograph of a fabricated chip (scale bar is 1 mm). }
    
\label{fig:1}
\end{figure*} 

Our results demonstrate the integration of coherent quantum emitters with programmable photonic circuitry in a GaAs platform, while providing a framework for independent electrical control, optical routing, and active circuit reconfiguration. 
This approach points toward a new class of III-V quantum photonic platforms where multiple quantum emitters are interconnected through programmable circuits to implement different quantum functionalities, such as photon routing, quantum interference or emitter-mediated nonlinear interactions.

\section{Programmable quantum photonic architecture}

The programmable quantum photonic platform presented in this work consists of several building blocks that have been developed individually over the past years, including a GaAs/InAs quantum-dot-based single-photon source \cite{uppu2020scalable}, shallow-etched grating couplers \cite{zhou2018high}, and a NEMS-based, electrostatically tunable waveguide-gap directional coupler, which serves as a nanomechanical router \cite{papon2019nanomechanical}.
Figure \ref{fig:1}a illustrates the key components. Quantum emitter–photon interfaces are realized using single-line-defect photonic crystal waveguides (PCWs). These structures provide enhanced light–matter interaction and Purcell enhancement, enabling efficient coupling between individual QDs and propagating optical modes \cite{Arcari2014Near-UnityWaveguide}. To enable independent operation of multiple emitters on the same chip, each PCW is electrically isolated from neighbouring devices using shallow trench structures \cite{Papon2023IndependentEmitters}. This architecture enables simultaneous control of spectrally distinct QDs while suppressing unwanted electrical crosstalk.

\begin{figure*}[!ht]
\centering
\includegraphics{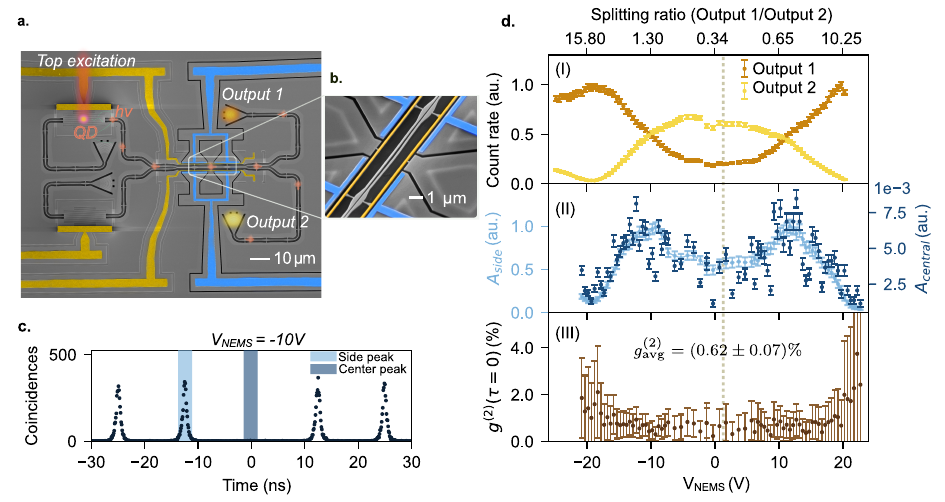}
\caption{\textbf{Operation of the nanomechanical router with single photons.} \textbf{(a.)} The QD is excited from the top, and the emitted photons are routed through the NEMS device and detected at the two output grating couplers using independent SNSPDs. \textbf{(b.)} Zoom-in of the NEMS switch. \textbf{(c.)} Second-order correlation histogram at $V_{\mathrm{NEMS}}=-10$~V, showing suppression of the zero-delay peak. \textbf{(d.)}(I) Voltage-dependent output count rates;(II) areas of the central ($A_\mathrm{center}$) and side ($A_\mathrm{side}$) peaks extracted from the second-order correlation measurements; (III) $g^{(2)}(0)=A_\mathrm{center}/A_\mathrm{side}$, showing high single-photon purity throughout the tuning range, with an average value of $(0.62\pm0.07)\%$. 
The upper $x$-axis shows the splitting ratio for each $V_{\mathrm{NEMS}}$, calculated as the ratio of Output 1 to Output 2 in (I).
The grey dashed line marks the QD bias voltage.
}
\label{fig:2}
\end{figure*}

Optical interconnects between different circuit functionalities are formed by suspended single-mode GaAs waveguides supported by nanoscale tethers. These waveguides exhibit propagation losses below 7~dB mm$^{-1}$ while supporting bending radii as small as 3~$\mu$m, enabling compact and densely integrated photonic circuits. For the circuit dimensions investigated here ($<100~\mu$m), the propagation losses are generally negligible. Light is coupled into and out of the chip using grating couplers enhanced by a distributed Bragg reflector, providing fiber-to-chip coupling losses below 2~dB per interface.\\
Reconfigurability is provided by NEMS, which implements low-loss, low-power, and cryogenic-compatible photonic actuators. The platform supports both tunable beam splitters and phase shifters, enabling programmable control of optical routing and interference with insertion losses below 1~dB, switching speeds in the MHz regime, and phase shifts exceeding $2\pi$ \cite{qvotrup2025integration}.\\
Electrical signal routing is achieved via a network of Cr/Au electrodes and wire-on-wire crossings, which enable independent control of multiple electrically-isolated components. This approach eliminates the need for complex multilayer metallization stacks and enables a simplified back-end-of-line process (see Supplementary Information for further details on the interconnect fabrication).\\
A central challenge in combining QDs and a nanomechanical router within a common circuit arises from their substantially different operating voltages. While QDs diodes are typically controlled in forward bias between 0 and 1.4 V, nanomechanical elements require actuation voltages of up to several tens of volts. To enable simultaneous operation, the NEMS electrode is electrically isolated from the n-doped contact layer and referenced instead to the p-i-n diode anode, as illustrated in Fig.~\ref{fig:1}b. This configuration establishes a common voltage reference while maintaining independent control of the optical and mechanical subsystems. Under normal operating conditions, QD devices exhibit conventional diode behavior, whereas nanomechanical routers behave as open circuits with leakage currents in the pA range (Fig.~\ref{fig:1}c). This isolation scheme forms a key element of the platform and enables the co-integration of multiple active photonic components without compromising the QD performance.\\
Figure \ref{fig:1}d summarizes the post-fabrication device layout used to realize the platform. The process combines all functionalities within a single suspended membrane architecture through a sequence of lithographic, etching, metallization and deposition steps. The final devices comprise parallel arrays of components electrically connected to common pads for wire bonding to printed circuit boards. A photo of a representative fabricated and wire-bonded chip is shown in Fig.~\ref{fig:1}e.\\
To demonstrate the coordinated operation of quantum emitters and reconfigurable photonic circuitry, we employ the circuit shown in Fig.~\ref{fig:2}a. The device comprises two independently controlled QD photon-emitter interfaces connected through a nanomechnical router (detail Fig.~\ref{fig:2}b). The implementation establishes the control and interconnect framework required for larger systems, where the primary limitations are expected to arise from cryogenic packaging and electrical/optical interfacing rather than from the underlying photonic architecture itself. Crucially, the ability to simultaneously operate multiple coherent emitters and a reconfigurable photonic element within a single architecture has not previously been demonstrated in a III--V quantum photonic platform. 

\section{Single-Photon routing}
\begin{figure*} [!ht]
    \centering
    \includegraphics[width=\textwidth]{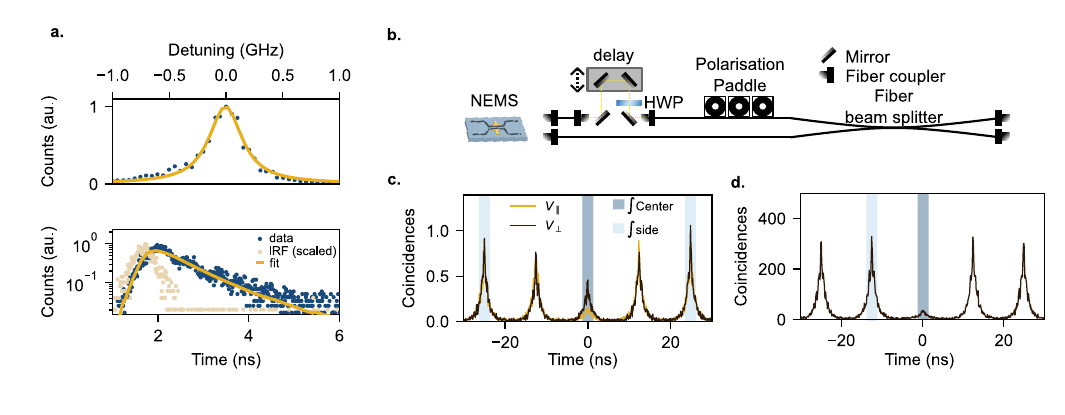}
    \caption{\textbf{Quantum dot single photon source integrated with nanomechanical router.} \textbf{(a.)} Upper panel: Resonance fluorescence spectrum of a single quantum-dot exciton transition measured under resonant continuous-wave excitation. A Voigt profile fit (orange) yields a full width at half maximum of $\Gamma_{\mathrm{RF}}=(351.8\pm28.0)~\mathrm{MHz}$. Lower panel: Time-resolved fluorescence under resonant $\pi$-pulse excitation. A single-exponential model convolved with the instrument response function (IRF) yields a decay rate of $\gamma=(1.55\pm0.08)~\mathrm{ns}^{-1}$, corresponding to a transform-limited linewidth of $246.2\pm13.3~\mathrm{MHz}$. \textbf{(b.)} Schematic of the two photon interference setup, in which NEMS serve as the first beam splitter. \textbf{(c.)}  Hong--Ou--Mandel (HOM) measurement for parallel (indistinguishable) and perpendicular (fully distinguishable) configurations, controlled by a half-wave plate (HWP) in the delay arm. The NEMS is tuned to a $50:50$ splitting ratio. Raw interference visibility of
    $V_{\mathrm{HOM}}=(A_{\perp}-A_{\parallel})/A_{\perp}=(62.2\pm6.7)\%$
    using an integration window of $4\gamma$. \textbf{(d.)} Hanbury Brown--Twiss autocorrelation measurement recorded from one output of the NEMS at the $50:50$ configuration. Extracted second-order correlation,
    $g^{(2)}(0)=A_{\mathrm{center}}/A_{\mathrm{side}}=(12.3\pm3.2)\%$. 
    }

    \label{fig3}    
\end{figure*}
To evaluate the performance of the nanomechanical router in the quantum regime, we initially demonstrate an on-chip Hanbury Brown and Twiss (HBT) measurement using single photons emitted from a single QD embedded in the PCW. 
The emitter is excited quasi-resonantly using an 80 MHz pulsed laser and operated under forward electrical bias to stabilize the local charge environment (see Methods), thereby suppressing spectral fluctuations and enabling highly pure and blinking-free single-photon generation \cite{pedersen2020near, lobl2017narrow}. Photons emitted into the waveguide are routed through the nanomechanical router and collected from the two output grating couplers. Residual excitation light is suppressed using narrowband spectral filters before the two output signals are detected by superconducting nanowire single-photon detectors (SNSPDs).\\
The splitting ratio of the on-chip beam splitter is controlled electrostatically through the NEMS actuator. We tune the device to the 50:50 operating point and observe strongly suppressed coincidence events at zero time delay as shown in the correlation histogram of Fig.~\ref{fig:2}c, yielding $g^{(2)}(0)=(0.7\pm0.5)\%$ consistent with near-ideal single-photon emission. The exceptionally low multi-photon probability reflects the high quality of the emitter.\\
To determine whether the single-photon character of the emitted light is preserved over different NEMS voltages and configurations, the HBT measurement is repeated while sweeping the NEMS voltage across its full tuning range. The corresponding output count rates are shown in Fig.~\ref{fig:2}d(I). As the NEMS voltage is varied, the relative photon flux in the two output ports changes continuously, demonstrating electrical control of the beam-splitter splitting ratio. A full $\pi$ phase shift of the mechanical response is observed at \(V_{\mathrm{NEMS}}=-18.5\)~V. The approximately symmetric response with respect to the voltage polarity is consistent with the electrostatic nature of the NEMS actuation.\\
The corresponding coincidence rate, given by the side-peak count rate in Fig.~\ref{fig:2}d(II), evolves consistently with the changing beam-splitter ratio and reaches its maximum near balanced 50:50 operation, where the probability of detecting photons in both output ports is highest. For clarity, the error bars on the side-peak and central-peak areas are scaled by factors of 10 and 0.1, respectively. In contrast, the normalized zero-delay coincidence rate remains approximately constant throughout the tuning range, yielding an average \(g^{(2)}(0)=(0.62\pm0.07)\%\), as shown in Fig.~\ref{fig:2}d(III). This demonstrates that active nanomechanical reconfiguration does not measurably affect the single-photon purity. The increased uncertainty near highly asymmetric splitting ratios results from reduced coincidence counts due to the lower photon flux in one output arm, rather than from a change in photon statistics of the emitted light. 

\section{Coherent Single Photon Emission under resonance fluorescence}

Besides enabling active routing and single-photon operation, the integrated architecture preserves the coherence required for quantum interference. The degree of coherence is determined by resonance-fluorescence (RF) spectroscopy of individual QD lines. A kHz-linewidth continuous-wave (cw) laser is scanned across an excitonic transition at an excitation power of $0.05~\mu\mathrm{W}$ and $V_{\mathrm{QD}}=1.303~\mathrm{V}$. We extract a RF linewidth of $\Gamma_{\mathrm{RF}} = 351.9 \pm 28.0~\mathrm{MHz}$, as shown in the upper panel of Fig.~\ref{fig3}a. We compare it to the natural linewidth of the emitter, obtained by time-resolved fluorescence under pulsed resonant excitation (lower panel of Fig.~\ref{fig3}a).
A single-exponential fit yields a lifetime of $T_1 = 0.646 \pm 0.035~\mathrm{ns}$ and a transform-limited linewidth of $(2\pi T_1)^{-1} = 246.2 \pm 13.3~\mathrm{MHz}$, implying that the RF linewidth is only a factor of $1.43 \pm 0.14$ broader than the transform limit, on par with previously reported values for gated QDs in PCWs \cite{pedersen2020near}.\\
Photon indistinguishability is investigated using a partially integrated Hong-Ou-Mandel (HOM) interferometer. As illustrated in Fig.~\ref{fig3}b, photons generated by a single QD are first crossing the on-chip NEMS router, which acts as the first beam splitter of the HOM interferometer and then are collected via two output gratings and coupled into separate optical fibers. One arm is delayed in an external free-space setup by one pulse period of the $80~\mathrm{MHz}$ excitation laser, allowing consecutively emitted photons to arrive simultaneously at a $50\!:\!50$ fiber beam splitter, where interference takes place. Polarization control in the delayed arm enables measurements using either parallel or orthogonal photon polarizations.\\
In the HOM experiment, the raw visibility is extracted comparing the integrated area of the central coincidence peak for orthogonal $A_\perp$ and parallel $A_\parallel$ configuration as
$V_{\mathrm{HOM,raw}}=\frac{A_{\perp}-A_{\parallel}}{A_{\perp}}$. Figure~\ref{fig3}c shows measured HOM in both polarization configurations.  In Figure~\ref{fig3}d a corresponding $g^{(2)}(0)$ measurement is shown. \\
The measured HOM visibility reaches $V_{\mathrm{HOM,raw}}=(62.2\pm6.7)\%$. Beyond the contribution of residual multiphoton events, quantified by the measured $g^{(2)}(0)$, experimental imperfections further reduce the observed visibility. In particular, residual resonant-laser background arising from the free-space objective-coupling scheme could be mitigated by direct fiber-to-grating coupling and integrated packaging, which would reduce sensitivity to polarization alignment and improve the suppression of residual excitation light.
Nevertheless, the near-transform-limited linewidth provides independent evidence of high optical coherence and, together with the HOM measurement, these results demonstrate that the coherence of the emitted photons is maintained during
active nanomechanical routing and subsequent interference within the programmable photonic circuit.\\

\begin{figure} [!ht]
\centering
\includegraphics{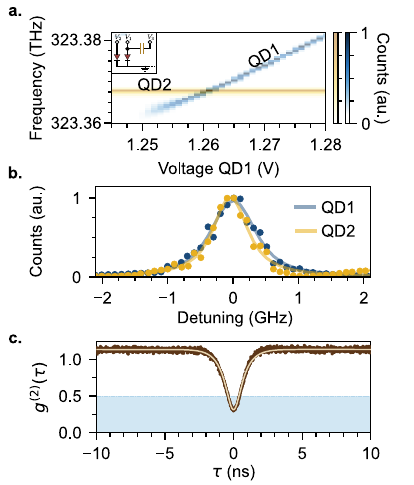}
\caption{\textbf{Simultaneous tunability of two independent quantum-dot sources.}
(\textbf{a.}) Collected emission from QD1, located in the upper source segment, as the bias voltage $V_{\mathrm{QD1}}$ is swept from $1.24$~V to $1.28$~V. During the sweep, QD2 in the lower source segment is biased at $V_{\mathrm{QD2}}=1.2265$~V while a narrowband continuous-wave (cw) laser is scanned across the QD2 resonance at $f_{\mathrm{res}}=323.367721$~THz.
(\textbf{b.}) Spectral overlap of the two quantum-dot transitions at the crossing frequency $f_{\mathrm{res}}$. The corresponding bias voltages are $V_{\mathrm{QD1}}=1.261$~V and $V_{\mathrm{QD2}}=1.2265$~V. Individual Voigt fits yield linewidths of $\Gamma_{\mathrm{QD1}}=(818\pm23)$~MHz and $\Gamma_{\mathrm{QD2}}=(632\pm56)$~MHz.
(\textbf{c.}) Second-order correlation measurement between photons detected at Output 1 and 2, while both QDs are simultaneously excited under cw resonant excitation, revealing non-classical two-photon interference.}
\label{fig:2QD}
\end{figure}

To demonstrate a step towards the scalability of the developed platform, we investigate the operation of two independently controlled single-photon sources. The p-i-n diode incorporated into the QD device layer enables electrical tuning of the emission frequency through the quantum-confined Stark effect, compensating for the inhomogeneous spectral distribution of QDs across the chip. This capability is an important prerequisite for the realization of multi-emitter quantum networks \cite{kuhlmann2015transform, Papon2023IndependentEmitters}. The integrated electrical interconnects and isolation trenches enable independent control of the two single-photon sources while preventing unwanted electrical coupling between their control electrodes.\\
Figure~\ref{fig:2QD}a demonstrates this functionality through a voltage--frequency map of two independently controlled QDs. QD1 (QD2) is located in the upper (lower) PCW and is controlled by an individual voltage source, \(V_1\) (\(V_2\)), through an electrically isolated contact, as illustrated in the inset of Fig.~\ref{fig:2QD}a. The map is obtained by resonantly exciting each QD individually and subsequently combining the corresponding RF spectra in post-processing. At the point where the two transitions are spectrally aligned, the RF spectra are shown in Fig.~\ref{fig:2QD}b, exhibiting linewidths of \((820\pm20)~\mathrm{MHz}\) and \((630\pm50)~\mathrm{MHz}\) for QD1 and QD2, respectively.

 \begin{figure*}[!ht]
    \centering
    \includegraphics{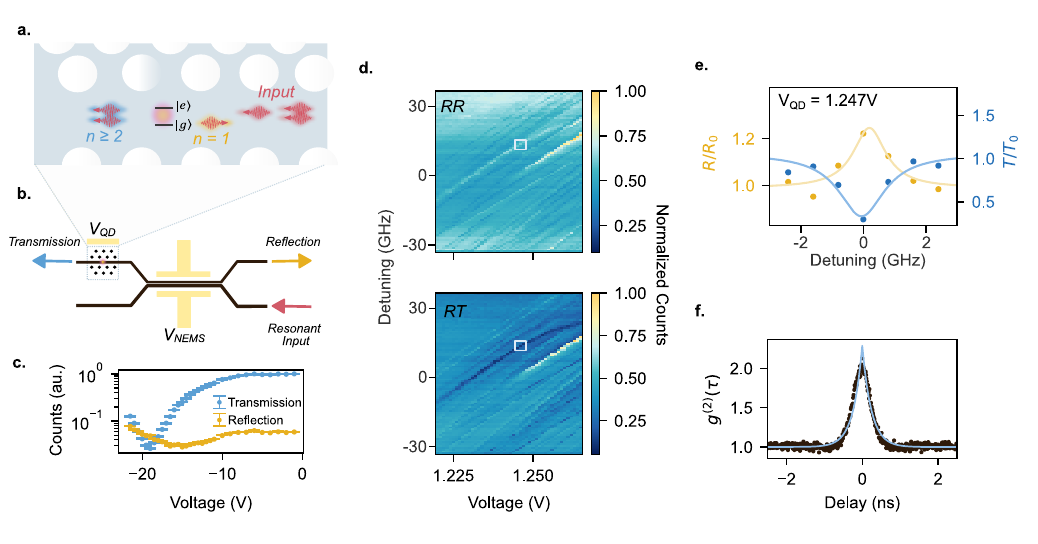}
\caption{\textbf{Nanomechanical routing for resonant scattering measurements.}
\textbf{(a.)} Schematic of single- and multi-photon scattering from a two-level emitter in a photonic crystal waveguide. Single photons are coherently reflected, whereas multi-photon components are predominantly transmitted due to emitter saturation \textbf{(b.)} Scheme of the resonant reflection (RR) and resonant transmission (RT) measurements using nanomechnical router. \textbf{(c.)} Reflection and transmission count rates versus NEMS actuation voltage, with minimum transmission leakage at $V_{\mathrm{NEMS}}=-19$ V.\textbf{(d.)} RR and RT maps recorded at $V_{\mathrm{NEMS}}=-19$ V as a function of excitation laser detuning and quantum-dot bias voltage.\textbf{(e.)} Representative RR and RT spectra showing enhanced reflection and extinction in transmission at the quantum-dot resonance. Solid lines show Lorentzian fits.\textbf{(f.)} Hanbury Brown--Twiss measurement in transmission showing photon bunching, with a correlation decay time of $\tau_b=(238.7\pm4.2)$ ps.
}
    \label{fig4}    

\end{figure*}

The integrated nanomechanical router further enables on-chip two-photon interference (TPI) measurements between the two QDs. Both emitters are resonantly driven under cw excitation, without spectral filtering, and the emitted photons interfere at the integrated beam splitter before being collected from its two output ports. A cross-correlation measurement between the outputs yields the $g^{(2)}(\tau)$ function shown in fig~\ref{fig:2QD}c. The pronounced suppression at zero time delay, reaching $g^{(2)}(0)=0.312\pm0.003$, lies below the classical limit of 0.5 expected for distinguishable photons and provides an evidence of quantum interference between the emitters. The residual background is primarily attributed to laser leakage, as discussed above, and is difficult to eliminate completely with free-space excitation and collection setups. Additional details of the TPI characterization are provided in the Supplementary Information. Nevertheless, these results demonstrate the potential of the platform for the simultaneous control of multiple emitters and the realization of on-chip quantum interference experiments.

\section{Quantum-Dot-Mediated Resonant Scattering}

Beyond single-photon routing, the nanomechanical router can also be used to control the nonlinear optical response of a QD. Resonant photon-scattering processes have potential applications in quantum photonic technologies, including photon sorting and Bell-state analysis \cite{ralph2015photon, witthaut2012photon, nielsen2026photon}. As a proof-of-principle application, Fig.~\ref{fig4}a illustrates the use of a single QD as a resonant scatterer in combination with NEMS routers.\\
A weak resonant laser coupled to the PCW coherently scatters from the QD \cite{le2021experimental, chang2014quantum, lodahl2015interfacing}, leading to a transmission dip and reflection peak at zero detuning. At higher powers, saturation of the two-level transition reduces the scattering response, resulting in photon-number-dependent transmission and reflection.
This provides a mechanism for nonlinear optical filtering and enables modification of the photon statistics of the transmitted field.\\
To read out both transmitted and reflected fields, as required, for instance, in Bell-state analysis protocols or photon-number sorters, we employ the NEMS router as shown in Fig.~\ref{fig4}b and tune its splitting ratio to the maximum-extinction operating point. This corresponds to a transmission extinction ratio of $15.9~\mathrm{dB}$ at $V_{NEMS} = -19~V$, enabling efficient separation of the reflected QD emission from the excitation laser.\\
Figure~\ref{fig4}c shows the measured transmission and reflection as a function of the NEMS voltage, with the laser frequency tuned to the resonance of the selected QD at 326.7686 THz. The measurement was performed with the QD turned off.
With the excitation laser coupled into one of the grating couplers (marked in red), light is simultaneously collected from both ends of the PCW, corresponding to the resonant transmission (RT) (blue) and resonant reflection (RR) (orange). Both the excitation laser frequency and the QD bias are scanned to measure the resonant response of the emitter. The resulting RT and RR maps are shown in Fig.~\ref{fig4}d. At resonance, the coherent interaction between the QDs and the guided optical mode produces a pronounced extinction in the transmission channel and an enhanced signal in the reflection channel. Representative transmission and reflection spectra, corresponding to the grey box in Fig.~\ref{fig4}d, are presented in Fig.~\ref{fig4}e. Lorentzian fits to the spectra yield visibilities of $(42.7 \pm 14.9)\%$ for transmission and $(11.8 \pm 4.8)\%$ for reflection. The measured visibilities are below the ideal values, which can be attributed in part to the finite frequency step size used during the laser scan, as well as to imperfect mode matching and residual background contributions.\\
Finally, to investigate the quantum nature of the transmitted field, photons collected in the transmission channel are sent to an off-chip HBT setup to measure second-order autocorrelation. A pronounced photon bunching of $g^{(2)}(0) = (2.291 \pm 0.010)$ is observed in the scattered light.  (Fig.~\ref{fig4}f). This behavior originates from the nonlinear interaction between the incident field and the QD which subtracts single-photon components from the weak coherent field \cite{le2021experimental}. \\
The combined functionality of tunable beam splitter and coherent photon-emitter interface extends the role of the NEMS device well beyond single photon routing, enabling on-chip photon-number based routing with adaptive behaviour.

\section{Conclusions}

We have demonstrated a programmable GaAs quantum photonic circuit integrating independently tunable QD emitters and NEMS routers operating at cryogenic temperatures. Our data showed consistency over four runs with in-between warm-up and cool-downs. Importantly, this was achieved while preserving coherent and non-classical light generation, enabling single-photon generation, two-photon interference, and resonant transmission and reflection measurements. \\
This work provides an enabling step toward larger-scale quantum photonic circuits. Further scaling will be enabled by continue advances in both optical and electrical interfacing. On the optical side, direct chip-to-fiber interfaces based on fiber arrays, multicore fibers, or advanced photonic packaging technologies will be required to efficiently address large numbers of optical ports and to reduce laser leakage, thereby directly improving measurements of HOM visibility under resonance fluorescence. On the electrical side, future circuits will require substantially larger numbers of control channels. Scaling beyond conventional wire-bonding approaches will likely require the co-integration of specially-designed cryogenic control electronics, including multiplexing, digital-to-analog conversion, and drivers located in close proximity to the photonic chip \cite{incandela2018characterization}.\\
The platform could be further expanded with the integration of on-chip SNSPDs \cite{sprengers2011waveguide, erbe2024mo} which would enable fully integrated quantum photonic circuits within a single material system. Heterogeneous integration approaches, including the transfer printing of electro-optic materials such as TFLN or barium titanate (BTO) onto GaAs photonic circuits \cite{mazet2015review}, could provide high-speed modulation and switching functionalities complementary to the low-loss nanomechanical elements employed here. In parallel, extending the platform towards telecom wavelengths around 1300 nm \cite{albrechtsen2026quantum} could combine low-loss circuits \cite{chang2019low}, native electro-optic functionality \cite{li2025engineering}, and high-quality quantum emitters within the same architecture.

\section{Methods}

\subsection*{Fabrication process}

The fabricated GaAs chip has dimensions of $4.5\times4.5~\mathrm{mm}^2$ and is based on an epitaxially grown heterostructure containing self-assembled InAs QDs embedded in a \textit{p--i--n} diode. The GaAs device layer has a thickness of 174~nm with a 1150~nm-thick $Al_{0.75}Ga_{0.25}As$ sacrificial layer below. A distributed Bragg reflector (DBR), consisting of 15 pairs of 79~nm AlAs and 66~nm GaAs layers, is incorporated beneath the sacrificial layer. The DBR reflects photons emitted towards the substrate back towards the surface, thereby enhancing the collection efficiency of the QD emission. During fabrication, the AlGaAs sacrificial layer is selectively removed using hydrofluoric (HF) acid to release the suspended photonic structures and NEMS components.\\
The QDs emit at wavelengths around 930~nm and provide the single-photon sources used in the photonic circuit. The photonic structures are fabricated directly in the GaAs device layer, enabling the integration of QDs with waveguides, grating couplers, photonic crystal structures, and NEMS devices on the same chip. The photonic structures are defined using electron-beam lithography followed by dry etching. Two etch depths are used to realize the different device functionalities. Shallow-etched structures, including the grating couplers and electrical isolation trenches around the QD sources, are etched through the p-layer to a depth of approximately 80~nm. The shallow trenches define the regions associated with the QD sources and electrically isolate their p-contacts from the surrounding semiconductor. Deeper structures, including the PCWs, cavities, NEMS structures, and electrical isolation trenches around the NEMS electrodes, are etched through the GaAs device layer down to the sacrificial AlGaAs layer, corresponding to an etch depth of approximately 180~nm. The deeper trench surrounding the NEMS electrical line provides electrical isolation from the surrounding doped semiconductor while also defining the regions that are subsequently released.\\
The electrical contacts are fabricated by electron-beam lithography followed by deposition of a 10~nm Cr / 70~nm Au metal stack. Chromium serves as an adhesion layer, while the Au layer provides the required electrical conductivity. The total metal thickness of approximately 80~nm is kept relatively small to limit mechanical loading of the suspended structures while maintaining sufficiently conductive electrodes for electrostatic actuation. The wire-bonding pads receive an additional 100~nm Au layer to improve the mechanical reliability of the wire-bonds when the chip is mounted on the external PCB.

\subsection*{Optical setup}

The fabricated chip is mounted inside a closed-cycle cryostat operating at approximately 5~K. The QD sources and the nanomechanical router are connected to independent external voltage sources, enabling separate electrical control of the QDs and the NEMS actuator.\\
Optical excitation and collection are performed through the top optical window of the cryostat using a 70$\times$ objective with a numerical aperture of NA = 0.81. The sample is mounted on a stack of three-axis nanopositioners (Attocube), allowing precise positioning of the chip with respect to the optical axis.\\
The optical setup provides two independent excitation paths, one for each QD source. The emitted light is collected through two orthogonally oriented shallow-etched grating couplers. The two collection gratings are spatially resolved in free-space and separated using a polarizing beamsplitter (PBS), with each grating designed to outcouple light with an orthogonal linear polarization. The cross-polarization configuration between excitation and collection further suppresses the reflected excitation laser, improving the signal-to-noise ratio of the detected single-photon signal.\\
The excitation lasers are transmitted to the optical breadboard through polarization-maintaining fibers and coupled into the setup using two fiber collimators. For top excitation, a collimator equipped with a lens of focal length $f = 18$~mm is used to produce a near-diffraction-limited spot on the sample surface for both excitation ports.\\
Each collection path is subsequently coupled into an optical fiber and directed to an individual channel of a SNSPD. The measurements are performed using PhotonSpot SNSPDs, with a specified timing jitter of 96~ps and an efficiency of approximately $70\%$. A time-tagger records the arrival time of each detection event, with a specified RMS timing jitter of 34~ps.\\
Two laser sources are primarily used for excitation throughout this work. For cw excitation, a tunable Toptica CTL laser is used, covering a wavelength range from approximately 900 to 960~nm with a specified linewidth below 10~kHz. For pulsed excitation, a tunable Picus laser is used to generate approximately 30~ps pulses at a repetition rate of 80~MHz. The broadband pulses are spectrally narrowed using an angle-dependent 15~GHz Alluxa optical filter before being coupled into the excitation path.\\

\noindent\textbf{Acknowledgements}
We acknowledge funding from the European
Research Council (ERC) under the European Union’s
Horizon 2020 research and innovation program (No.
949043, NANOMEQ), the Danish National Research Foundation (Center of Excellence “Hy-Q,” grant number
DNRF139), BMFTR QRN project
16KIS2200, as well as DFG excellence cluster ML4Q
project EXC 2004/1.\\
\noindent\textbf{Notes}
The authors declare no competing financial interest.\\
\noindent\textbf{Conflicts of interest}
P.L. is a founder of the company
Sparrow Quantum, which commercialises single-photon
sources. The remaining authors declare no other con-
flicts of interest.


\bibliography{Main.bib}

@article{carolan2015universal,
  title={Universal linear optics},
  author={Carolan, Jacques and Harrold, Christopher and Sparrow, Chris and Mart{\'\i}n-L{\'o}pez, Enrique and Russell, Nicholas J and Silverstone, Joshua W and Shadbolt, Peter J and Matsuda, Nobuyuki and Oguma, Manabu and Itoh, Mikitaka and others},
  journal={Science},
  volume={349},
  number={6249},
  pages={711--716},
  year={2015},
  publisher={American Association for the Advancement of Science}
}

@article{bogaerts2020programmable,
  title={Programmable photonic circuits},
  author={Bogaerts, Wim and P{\'e}rez, Daniel and Capmany, Jos{\'e} and Miller, David AB and Poon, Joyce and Englund, Dirk and Morichetti, Francesco and Melloni, Andrea},
  journal={Nature},
  volume={586},
  number={7828},
  pages={207--216},
  year={2020},
  publisher={Nature Publishing Group UK London}
}

@article{paesani2019generation,
  title={Generation and sampling of quantum states of light in a silicon chip},
  author={Paesani, Stefano and Ding, Yunhong and Santagati, Raffaele and Chakhmakhchyan, Levon and Vigliar, Caterina and Rottwitt, Karsten and Oxenl{\o}we, Leif K and Wang, Jianwei and Thompson, Mark G and Laing, Anthony},
  journal={Nature Physics},
  volume={15},
  number={9},
  pages={925--929},
  year={2019},
  publisher={Nature Publishing Group UK London}
}

@article{wang2018integrated,
  title={Integrated lithium niobate electro-optic modulators operating at CMOS-compatible voltages},
  author={Wang, Cheng and Zhang, Mian and Chen, Xi and Bertrand, Maxime and Shams-Ansari, Amirhassan and Chandrasekhar, Sethumadhavan and Winzer, Peter and Lon{\v{c}}ar, Marko},
  journal={Nature},
  volume={562},
  number={7725},
  pages={101--104},
  year={2018},
  publisher={Nature Publishing Group UK London}
}

@article{zhu2021integrated,
  title={Integrated photonics on thin-film lithium niobate},
  author={Zhu, Di and Shao, Linbo and Yu, Mengjie and Cheng, Rebecca and Desiatov, Boris and Xin, CJ and Hu, Yaowen and Holzgrafe, Jeffrey and Ghosh, Soumya and Shams-Ansari, Amirhassan and others},
  journal={Advances in Optics and Photonics},
  volume={13},
  number={2},
  pages={242--352},
  year={2021},
  publisher={Optical Society of America}
}

@article{errando2019mems,
  title={MEMS for photonic integrated circuits},
  author={Errando-Herranz, Carlos and Takabayashi, Alain Yuji and Edinger, Pierre and Sattari, Hamed and Gylfason, Kristinn B and Quack, Niels},
  journal={IEEE Journal of Selected Topics in Quantum Electronics},
  volume={26},
  number={2},
  pages={1--16},
  year={2019},
  publisher={IEEE}
}

@article{larocque2024tunable,
  title={Tunable quantum emitters on large-scale foundry silicon photonics},
  author={Larocque, Hugo and Buyukkaya, Mustafa Atabey and Errando-Herranz, Carlos and Papon, Camille and Harper, Samuel and Tao, Max and Carolan, Jacques and Lee, Chang-Min and Richardson, Christopher JK and Leake, Gerald L and others},
  journal={Nature Communications},
  volume={15},
  number={1},
  pages={5781},
  year={2024},
  publisher={Nature Publishing Group UK London}
}

@article{qvotrup2025integration,
  title={Integration of a Ga As-based nanomechanical phase shifter with quantum-dot single-photon sources},
  author={Qvotrup, Celeste and Wang, Ying and Albrechtsen, Marcus and Thomas, Rodrigo A and Liu, Zhe and Scholz, Sven and Ludwig, Arne and Midolo, Leonardo},
  journal={Physical Review Applied},
  volume={24},
  number={5},
  pages={054016},
  year={2025},
  publisher={APS}
}

@article{papon2019nanomechanical,
  title={Nanomechanical single-photon routing},
  author={Papon, Camille and Zhou, Xiaoyan and Thyrrestrup, Henri and Liu, Zhe and Stobbe, S{\o}ren and Schott, R{\"u}diger and Wieck, Andreas D and Ludwig, Arne and Lodahl, Peter and Midolo, Leonardo},
  journal={Optica},
  volume={6},
  number={4},
  pages={524--530},
  year={2019},
  publisher={Optical Society of America}
}

@article{uppu2020scalable,
  title={Scalable integrated single-photon source},
  author={Uppu, Ravitej and Pedersen, Freja T and Wang, Ying and Olesen, Cecilie T and Papon, Camille and Zhou, Xiaoyan and Midolo, Leonardo and Scholz, Sven and Wieck, Andreas D and Ludwig, Arne and others},
  journal={Science advances},
  volume={6},
  number={50},
  pages={eabc8268},
  year={2020},
  publisher={American Association for the Advancement of Science}
}

@article{tomm2021bright,
  title={A bright and fast source of coherent single photons},
  author={Tomm, Natasha and Javadi, Alisa and Antoniadis, Nadia Olympia and Najer, Daniel and L{\"o}bl, Matthias Christian and Korsch, Alexander Rolf and Schott, R{\"u}diger and Valentin, Sascha Ren{\'e} and Wieck, Andreas Dirk and Ludwig, Arne and others},
  journal={Nature Nanotechnology},
  volume={16},
  number={4},
  pages={399--403},
  year={2021},
  publisher={Nature Publishing Group UK London}
}

@article{lobl2017narrow,
  title={Narrow optical linewidths and spin pumping on charge-tunable, close-to-surface self-assembled quantum dots in an ultra-thin diode},
  author={L{\"o}bl, Matthias C and S{\"o}llner, Immo and Javadi, Alisa and Pregnolato, Tommaso and Schott, R{\"u}diger and Midolo, Leonardo and Kuhlmann, Andreas V and Stobbe, S{\o}ren and Wieck, Andreas D and Lodahl, Peter and others},
  journal={arXiv preprint arXiv:1708.00520},
  year={2017}
}

@article{kuhlmann2015transform,
  title={Transform-limited single photons from a single quantum dot},
  author={Kuhlmann, Andreas V and Prechtel, Jonathan H and Houel, Julien and Ludwig, Arne and Reuter, Dirk and Wieck, Andreas D and Warburton, Richard J},
  journal={Nature communications},
  volume={6},
  number={1},
  pages={8204},
  year={2015},
  publisher={Nature Publishing Group UK London}
}

@article{silverstone2014chip,
  title={On-chip quantum interference between silicon photon-pair sources},
  author={Silverstone, Joshua W and Bonneau, Damien and Ohira, Kazuya and Suzuki, Nob and Yoshida, Haruhiko and Iizuka, Norio and Ezaki, Mizunori and Natarajan, Chandra M and Tanner, Michael G and Hadfield, Robert H and others},
  journal={Nature Photonics},
  volume={8},
  number={2},
  pages={104--108},
  year={2014},
  publisher={Nature Publishing Group}
}

@article{kuttner2026heralded,
  title={Heralded quantum interference in integrated lithium niobate nanophotonics},
  author={Kuttner, Tristan and Sabatti, Alessandra and Kellner, Jost and Grange, Rachel and Chapman, Robert J},
  journal={Optica Quantum},
  volume={4},
  number={2},
  pages={121--129},
  year={2026},
  publisher={Optica Publishing Group}
}

@article{politi2008silica,
  title={Silica-on-silicon waveguide quantum circuits},
  author={Politi, Alberto and Cryan, Martin J and Rarity, John G and Yu, Siyuan and O'brien, Jeremy L},
  journal={Science},
  volume={320},
  number={5876},
  pages={646--649},
  year={2008},
  publisher={American Association for the Advancement of Science}
}

@article{crespi2013integrated,
  title={Integrated multimode interferometers with arbitrary designs for photonic boson sampling},
  author={Crespi, Andrea and Osellame, Roberto and Ramponi, Roberta and Brod, Daniel J and Galvao, Ernesto F and Spagnolo, Nicolo and Vitelli, Chiara and Maiorino, Enrico and Mataloni, Paolo and Sciarrino, Fabio},
  journal={Nature photonics},
  volume={7},
  number={7},
  pages={545--549},
  year={2013},
  publisher={Nature Publishing Group UK London}
}

@article{midolo2012spontaneous,
  title={Spontaneous emission control of single quantum dots by electromechanical tuning of a photonic crystal cavity},
  author={Midolo, L and Pagliano, F and Hoang, TB and Xia, T and Van Otten, FWM and Li, LH and Linfield, EH and Lermer, M and H{\"o}fling, Sven and Fiore, A},
  journal={Applied Physics Letters},
  volume={101},
  number={9},
  year={2012},
  publisher={AIP Publishing}
}

@article{bishop2018electro,
  title={Electro-mechanical control of an on-chip optical beam splitter containing an embedded quantum emitter},
  author={Bishop, ZK and Foster, AP and Royall, B and Bentham, C and Clarke, E and Skolnick, MS and Wilson, LR},
  journal={Optics letters},
  volume={43},
  number={9},
  pages={2142--2145},
  year={2018},
  publisher={Optical Society of America}
}

@article{le2021experimental,
  title={Experimental reconstruction of the few-photon nonlinear scattering matrix from a single quantum dot in a nanophotonic waveguide},
  author={Le Jeannic, Hanna and Ramos, Tom{\'a}s and Simonsen, Signe F and Pregnolato, Tommaso and Liu, Zhe and Schott, R{\"u}diger and Wieck, Andreas D and Ludwig, Arne and Rotenberg, Nir and Garc{\'\i}a-Ripoll, Juan Jos{\'e} and others},
  journal={Physical Review Letters},
  volume={126},
  number={2},
  pages={023603},
  year={2021},
  publisher={APS}
}

@article{lodahl2015interfacing,
  title={Interfacing single photons and single quantum dots with photonic nanostructures},
  author={Lodahl, Peter and Mahmoodian, Sahand and Stobbe, S{\o}ren},
  journal={Reviews of Modern Physics},
  volume={87},
  number={2},
  pages={347--400},
  year={2015},
  publisher={APS}
}

@article{javadi2015single,
  title={Single-photon non-linear optics with a quantum dot in a waveguide},
  author={Javadi, Alisa and S{\"o}llner, I and Arcari, Marta and Hansen, S Lindskov and Midolo, Leonardo and Mahmoodian, Sahand and Kir{\v{s}}ansk{\.e}, G and Pregnolato, Tommaso and Lee, EH and Song, JD and others},
  journal={Nature communications},
  volume={6},
  number={1},
  pages={8655},
  year={2015},
  publisher={Nature Publishing Group UK London}
}

@article{pedersen2020near,
  title={Near transform-limited quantum dot linewidths in a broadband photonic crystal waveguide},
  author={Pedersen, Freja T and Wang, Ying and Olesen, Cecilie T and Scholz, Sven and Wieck, Andreas D and Ludwig, Arne and LObl, Matthias C and Warburton, Richard J and Midolo, Leonardo and Uppu, Ravitej and others},
  journal={ACS Photonics},
  volume={7},
  number={9},
  pages={2343--2349},
  year={2020},
  publisher={ACS Publications}
}

@article{Arcari2014Near-UnityWaveguide,
    title = {{Near-Unity Coupling Efficiency of a Quantum Emitter to a Photonic Crystal Waveguide}},
    year = {2014},
    journal = {Physical Review Letters},
    author = {Arcari, M. and S{\"{o}}llner, I. and Javadi, A. and Lindskov Hansen, S. and Mahmoodian, S. and Liu, J. and Thyrrestrup, H. and Lee, E. H. and Song, J. D. and Stobbe, S. and Lodahl, P.},
    number = {9},
    month = {8},
    volume = {113},
    publisher = {American Physical Society},
    doi = {10.1103/PhysRevLett.113.093603},
    issn = {10797114},
    arxivId = {1402.2081}
}

@article{Papon2023IndependentEmitters,
    title = {{Independent Operation of Two Waveguide-Integrated Quantum Emitters}},
    year = {2023},
    journal = {Physical Review Applied},
    author = {Papon, C. and Wang, Y. and Uppu, R. and Scholz, S. and Wieck, A. D. and Ludwig, A. and Lodahl, P. and Midolo, L.},
    number = {6},
    month = {6},
    volume = {19},
    publisher = {American Physical Society},
    doi = {10.1103/PhysRevApplied.19.L061003},
    issn = {23317019},
    arxivId = {2210.09826}
}

@article{zhou2018high,
  title={High-efficiency shallow-etched grating on GaAs membranes for quantum photonic applications},
  author={Zhou, Xiaoyan and Kulkova, Irina and Lund-Hansen, Toke and Hansen, Sofie Lindskov and Lodahl, Peter and Midolo, Leonardo},
  journal={Applied Physics Letters},
  volume={113},
  number={25},
  year={2018},
  publisher={AIP Publishing}
}

@article{nielsen2026photon,
  title={Photon Sorting with a Quantum Emitter},
  author={Nielsen, Kasper H and Corminboeuf, Etienne and Tissot, Benedikt and Pettersson, Love A and Scholz, Sven and Ludwig, Arne and Midolo, Leonardo and S{\o}rensen, Anders S and Lodahl, Peter and Wang, Ying and others},
  journal={arXiv preprint arXiv:2604.21758},
  year={2026}
}

@article{witthaut2012photon,
  title={Photon sorters and QND detectors using single photon emitters},
  author={Witthaut, Dirk and Lukin, Mikhail D and S{\o}rensen, Anders S},
  journal={EPL (Europhysics Letters)},
  volume={97},
  number={5},
  pages={50007},
  year={2012}
}

@article{ralph2015photon,
  title={Photon sorting, efficient Bell measurements, and a deterministic controlled-Z gate using a passive two-level nonlinearity},
  author={Ralph, TC and S{\"o}llner, I and Mahmoodian, S and White, AG and Lodahl, P},
  journal={Physical review letters},
  volume={114},
  number={17},
  pages={173603},
  year={2015},
  publisher={APS}
}

@article{Buzzi2025SpectralStrain,
    title = {{Spectral tuning and nanoscale localization of single color centers in silicon via controllable strain}},
    year = {2025},
    journal = {Nature Communications },
    author = {Buzzi, Alessandro and Papon, Camille and Pirro, Matteo and Hooybergs, Odiel and Raniwala, Hamza and Saggio, Valeria and Errando-Herranz, Carlos and Englund, Dirk},
    number = {1},
    month = {12},
    volume = {16},
    publisher = {Nature Research},
    doi = {10.1038/s41467-025-63871-8},
    issn = {20411723},
    pmid = {41044067},
    arxivId = {2501.17290}
}

@article{midolo2018nano,
  title={Nano-opto-electro-mechanical systems},
  author={Midolo, Leonardo and Schliesser, Albert and Fiore, Andrea},
  journal={Nature nanotechnology},
  volume={13},
  number={1},
  pages={11--18},
  year={2018},
  publisher={Nature Publishing Group UK London}
}

@article{chang2014quantum,
  title={Quantum nonlinear optics—photon by photon},
  author={Chang, Darrick E and Vuleti{\'c}, Vladan and Lukin, Mikhail D},
  journal={Nature Photonics},
  volume={8},
  number={9},
  pages={685--694},
  year={2014},
  publisher={Nature Publishing Group UK London}
}

@article{incandela2018characterization,
  title={Characterization and compact modeling of nanometer CMOS transistors at deep-cryogenic temperatures},
  author={Incandela, Rosario M and Song, Lin and Homulle, Harald and Charbon, Edoardo and Vladimirescu, Andrei and Sebastiano, Fabio},
  journal={IEEE Journal of the Electron Devices Society},
  volume={6},
  pages={996--1006},
  year={2018},
  publisher={IEEE}
}

@article{erbe2024mo,
  title={Mo-Si superconducting nanowire single-photon detectors on Ga As},
  author={Erbe, M and Berrazouane, R and Geyer, S and Stasi, L and van der Brugge, F and Gras, G and Schmidt, M and Wieck, AD and Ludwig, A and Bussi{\`e}res, F and others},
  journal={Physical Review Applied},
  volume={22},
  number={1},
  pages={014072},
  year={2024},
  publisher={APS}
}

@article{sprengers2011waveguide,
  title={Waveguide superconducting single-photon detectors for integrated quantum photonic circuits},
  author={Sprengers, JP and Gaggero, A and Sahin, D and Jahanmirinejad, S and Frucci, G and Mattioli, F and Leoni, R and Beetz, Johannes and Lermer, M and Kamp, M and others},
  journal={Applied Physics Letters},
  volume={99},
  number={18},
  year={2011},
  publisher={AIP Publishing}
}

@article{mazet2015review,
  title={A review of molecular beam epitaxy of ferroelectric BaTiO3 films on Si, Ge and GaAs substrates and their applications},
  author={Mazet, Lucie and Yang, Sang Mo and Kalinin, Sergei V and Schamm-Chardon, Sylvie and Dubourdieu, Catherine},
  journal={Science and technology of advanced materials},
  volume={16},
  number={3},
  pages={036005},
  year={2015},
  publisher={IOP Publishing}
}

@article{albrechtsen2026quantum,
  title={A quantum-coherent photon--emitter interface in the original telecom band},
  author={Albrechtsen, Marcus and Kr{\"u}ger, Severin and Loredo, Juan C and Stefan, Lucio and Liu, Zhe and Meng, Yu and Niekamp, Lukas L and Seyschab, Bianca F and Spitzer, Nikolai and Warburton, Richard J and others},
  journal={Nature Nanotechnology},
  pages={1--6},
  year={2026},
  publisher={Nature Publishing Group UK London}
}

@article{li2025engineering,
  title={Engineering cm-scale true push-pull electro-optic modulators in a suspended GaAs photonic integrated circuit platform by exploiting the orientation induced asymmetry of the Pockels r 41 coefficient},
  author={Li, Haoyang and Thomas, Robert and Jiang, Pisu and Balram, Krishna C},
  journal={Nanophotonics},
  volume={14},
  number={18},
  pages={3033--3042},
  year={2025},
  publisher={De Gruyter}
}

@article{chang2019low,
  title={Low loss (Al)GaAs on an insulator waveguide platform},
  author={Chang, Lin and Boes, Andreas and Pintus, Paolo and Xie, Weiqiang
          and Peters, Jon D. and Kennedy, M. J. and Jin, Warren and Guo, Xiao-Wen
          and Yu, Su-Peng and Papp, Scott B. and Bowers, John E.},
  journal={Optics Letters},
  volume={44},
  number={16},
  pages={4075--4078},
  year={2019},
  doi={10.1364/OL.44.004075}
}

@article{loredo2026deterministic,
  title={Deterministic quantum dot single-photon sources: Operational principles and state-of-the-art specifications},
  author={Loredo, JC and Stefan, L and Krogh, B and Jensen, R and Suleiman, I and Kr{\"u}ger, S and Bergamin, M and Thyrrestrup, H and Budtz, S and Roulund, J and others},
  journal={Applied Physics Reviews},
  volume={13},
  number={1},
  year={2026},
  publisher={AIP Publishing}
}

@article{meng2024deterministic,
  title={Deterministic photon source of genuine three-qubit entanglement},
  author={Meng, Yijian and Chan, Ming Lai and Nielsen, Rasmus B and Appel, Martin H and Liu, Zhe and Wang, Ying and Bart, Nikolai and Wieck, Andreas D and Ludwig, Arne and Midolo, Leonardo and others},
  journal={Nature communications},
  volume={15},
  number={1},
  pages={7774},
  year={2024},
  publisher={Nature Publishing Group UK London}
}
\end{document}